\documentclass[12pt,aps,floatfix,nofootinbib]{revtex4}
\usepackage{amsmath}
\usepackage{latexsym}
\usepackage{float}
\usepackage{amssymb}
\usepackage{graphicx}
\usepackage{epsfig}
\newcommand{\theeq}{\theta_\mathrm{eq}}
\newcommand{\thetr}{\theta^\infty_\mathrm{th}}

\begin{document}

\title{Capillary filling 
in microchannels patterned by posts }

\author{B.\ M.\ Mognetti,  J.\ M.\ Yeomans \\
The Rudolf Peierls Centre for Theoretical Physics,
 1 Keble Road Oxford, OX1 3NP, United Kingdom.
}

\begin{abstract}
We investigate the capillary filling of three dimensional micro-channels with 
surfaces patterned by posts of square cross section. We show that pinning on
the edges of the posts suppresses, and can halt, capillary filling. We stress 
the importance of the channel walls in controlling whether filling can occur. 
In particular for channels higher than the distance between adjacent posts, 
filling occurs for contact angles less than a threshold angle 
$\sim 55^\circ$, independent of the height of the channel.

\end{abstract}

\maketitle

% PACS numbers: 05.70.Ce, 64.70.F-, 64.75.Cd, 02.70.Tt
%(a) Electronic mail: b.mognetti1@physics.ox.ac.uk, j.yeomans1@physics.ox.ac.uk

\section{Introduction}

Capillary filling, the ability of water to fill a hydrophilic channel, has 
been recognised since  the pioneering work of Lucas and Washburn 
\cite{Washburn21,Lucas18,Bell1906} nearly a century ago. However, 
investigations of 
capillary filling in microchannels remain interesting due to modern 
applications of microfluid devices. Advances in lithographic techniques mean 
that it is becoming increasingly feasible to fabricate microchannels with well
defined surface structures on micron length scales. These have potential 
applications  for chemical detection \cite{detection}, as microreactors 
\cite{mreactor}, or to build entropic traps for DNA separation \cite{DNAsep}. 
Our aim in this paper is to present a numerical
investigation of how posts on the surface of a microchannel affect capillary 
filling. Our results are relevant to the use of electrowetting to control flow 
in microchannels and suggest ways to overcome the difficulties of filling 
structured microchannels.

If a channel with hydrophilic walls comes into contact with a fluid
reservoir it starts to fill as capillary forces pull the fluid into
the channel.  Balancing the capillary forces 
($2 \gamma \cos \theta_\mathrm{ad}$) against the viscous drag of the 
entering fluid ($12 \eta x (\mathrm{d}x/\mathrm{d} t) /H^2$)
gives an expression for the position of the advancing fluid
 in the channel $x$ as a function of time $t$ \cite{Washburn21},
\begin{equation}
x^2 = {\gamma H \cos \theta_\mathrm{ad}\over 3 \eta} \cdot t \, ,
\label{WashburnEq}
\end{equation}
for a capillary of height $H$ and infinite width. $\gamma$ is the 
gas-liquid surface tension, $\eta$ the fluid viscosity and $\theta_\mathrm{ad}$ 
the contact angle of the advancing front. This formula neglects inertial and 
gravitational effects (good approximations once filling is established, and for
 channels of dimension smaller than the capillary length), assumes
that the displaced fluid has zero viscosity, and neglects the slip length.
 These conditions are not always satisfied by simplified models 
used to investigation of capillary filling. However, excellent 
agreement between the theory and numerical results can be achieved 
by accounting for drag forces of the gas ($12 \eta_\mathrm{gas} 
(L-x) (\mathrm{d}x/\mathrm{d} t) /H^2$, where $L$ is the length of the 
channel and $\eta_\mathrm{gas}$ the viscosity of the gas)
\cite{CondEv}, or allowing for a slip length \cite{DMB-07}.

\begin{figure}[h]
\includegraphics[angle=-90,scale=0.28]{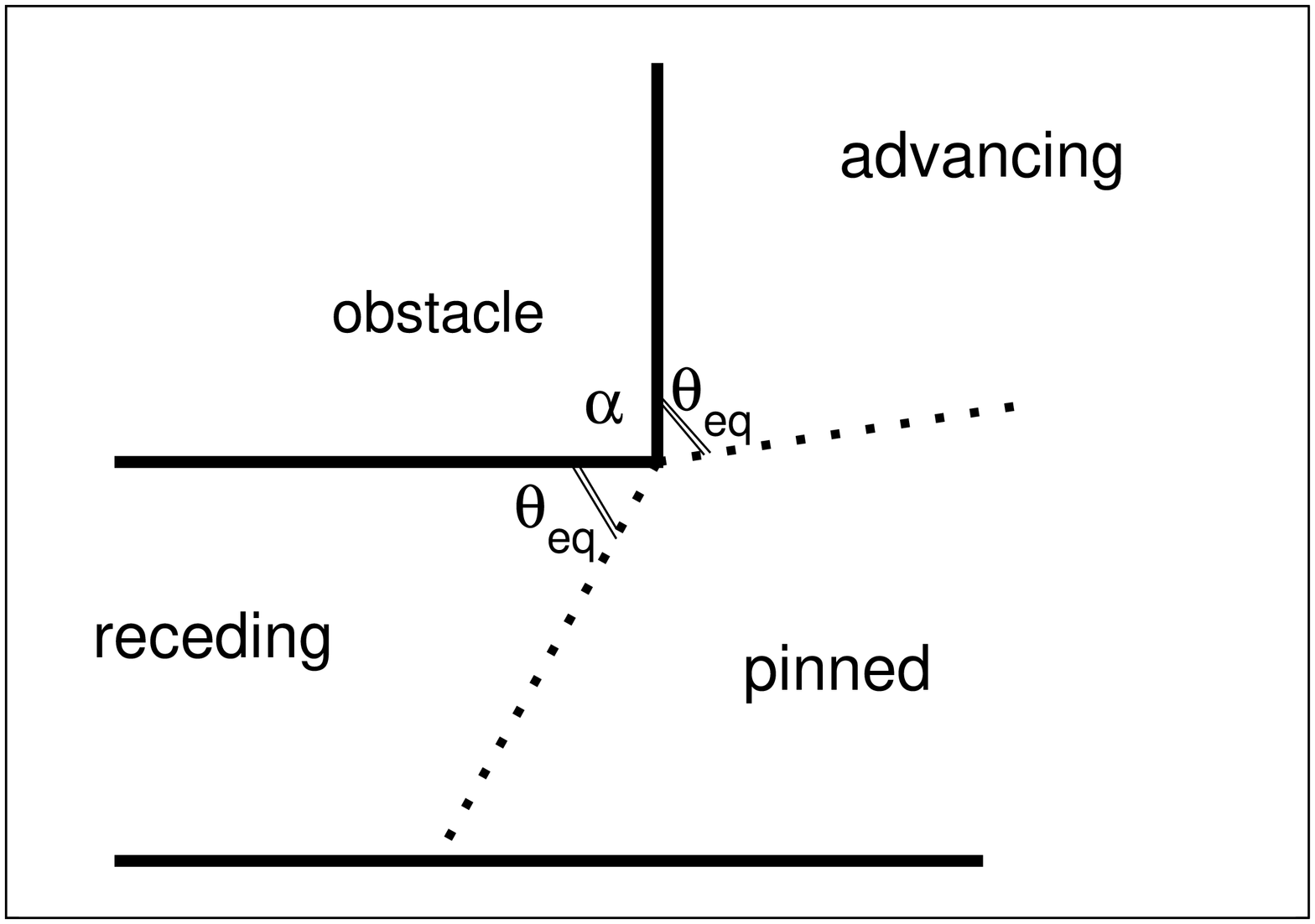}
\includegraphics[angle=-90,scale=0.28]{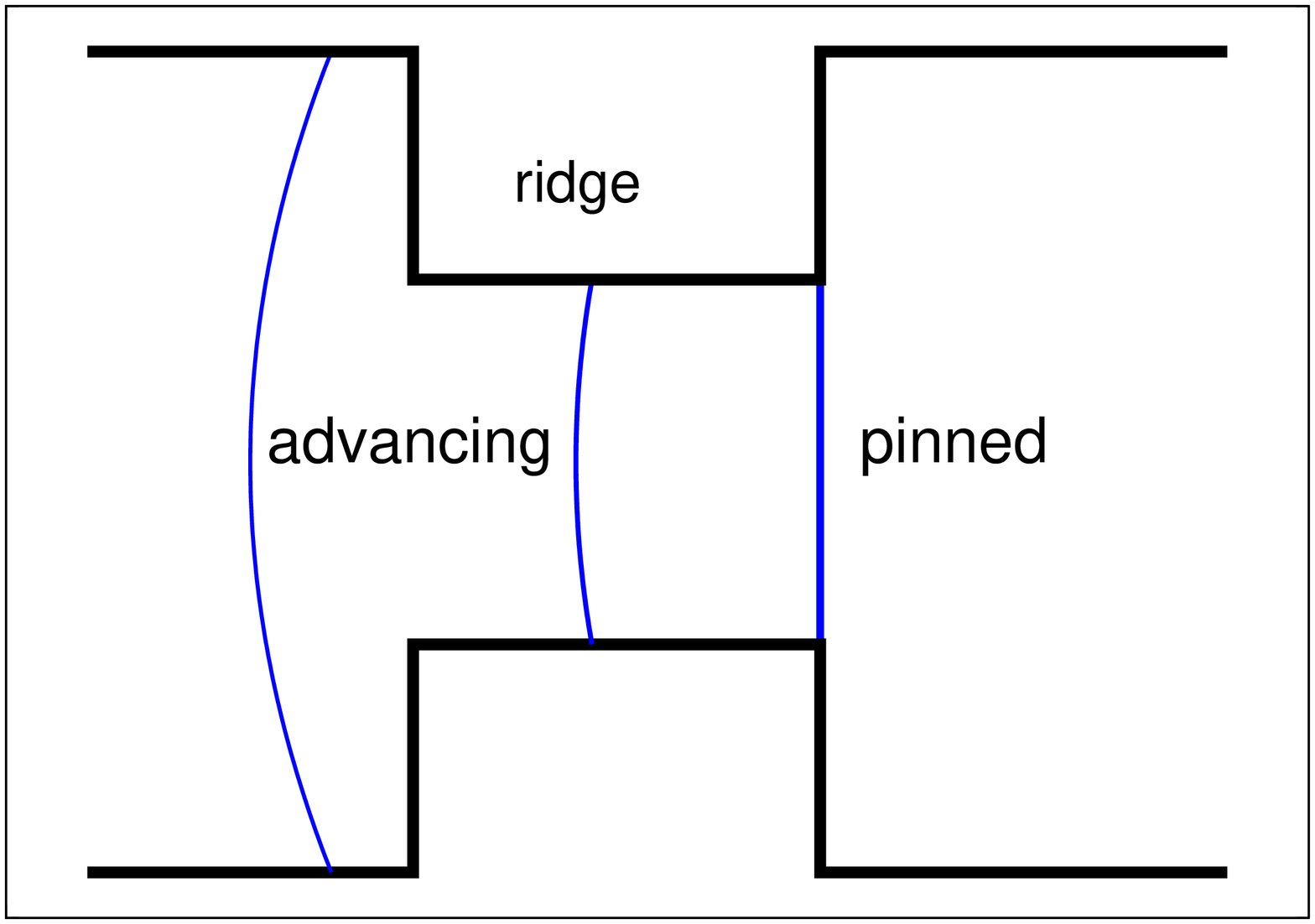}\\
(a) $\qquad\qquad\qquad\qquad\qquad\qquad\qquad\qquad\qquad\quad$ (b)
\caption{(Color online) (a) When a fluid interface reaches an edge
it remain pinned for a range of angles $2\pi-2 \theeq-\alpha$. 
(b) A fluid front, advancing from the left, remains pinned (straight line) at 
the edges of two opposing ridges for all contact angles.}\label{GibbsFig}
\end{figure}
When the surface of a microchannel is patterned with obstacles, such as posts 
or ridges, the capillary filling is, in general, suppressed because of pinning 
on the edges of the posts. The statics of the pinning can be understood by 
reference to the Gibbs' criterion \cite{Gibbs}. This states that, when a fluid 
interface reaches a edge, it will remain pinned over the range of angles 
between the equilibrium contact angle on each of the surfaces bounding the 
 edge as illustrated in Fig.~\ref{GibbsFig}(a). A striking consequence of the
 Gibbs' criterion is that if there are opposing ridges across the channel it 
does not fill \cite{Kusumaatmaja08}. Capillary forces pull the interface until 
it becomes a flat surface, pinned on the edges as shown in 
Fig.~\ref{GibbsFig}(b). It cannot move down the sides of the post from this 
configuration, and hence will remain pinned. If the meniscus is moving, 
however, inertial effects may allow it to 
overcome the pinning as it reaches the ridge. Then, as the interface moves 
down the channel it slows, and will eventually pin at a subsequent ridge.
We also note that for nanoscale roughness thermal fluctuations allow the 
interface to depin and advance \cite{MD}.

If the ridges across the channel are replaced by separated posts capillary 
filling becomes possible for sufficiently low values of the contact angle.
The aim of this paper is to investigate how the depinning behaviour depends on 
the channel geometry and the contact angle, and to explore the depinning 
mechanism in some detail. We emphasise the importance of the channel walls in 
controlling the depinning of the interface and we identify the way in which the
 interface depins for different channel geometries. Estimates are given for the
 contact angles at which depinning occurs.

In Sec.\ \ref{model} we define the model we use, summarising the equilibrium
properties and the equations of motion. We then describe the geometry of the 
channel and list the simulation parameters of the model. In Sec.\ \ref{secRes} 
we present our results. For two typical channels, using values of $\theeq$ 
which do not pin, we compare the filling rate to the similar case of a smooth
channel (without obstacles), showing how posts on the channel surface
slow the filling. We then concentrate on the filling/pinned transition.
The case in which the posts span the channel and that in which the posts do 
not meet across the channel are investigated in Secs.\ \ref{secHeff0} and 
\ref{secHeff} respectively. We find that, for high channels and long posts, 
the pinning-depinning threshold angle $\thetr \sim 55^\circ$. 
The depinning is driven by the walls and is independent of the 
channel geometry. For narrow channels or short posts, filling is possible for 
higher contact angles. In Sec.\ \ref{secIn} we discuss  the effect of inertia 
on the determination of the threshold contact angle $\thetr$. Finally Sec.\ 
\ref{secDis} presents our conclusions.

\section{The model}\label{model}

As we are considering micron length scales it is appropriate to describe the 
system using a mesoscale modeling approach. We choose to use a diffuse 
interface model, solved using a lattice Boltzmann algorithm \cite{Succi,
Yeomans-Leuven,LB1}, 
which has proven to be a useful tool to model the dynamics of 
fluids with moving interfaces. 
Of particular relevance here Ref.\ \cite{Pooley08} 
 demonstrates how this approach can be used to describe  capillary 
filling in smooth microchannels. We now give details of the 
model, the channel geometry and the simulations parameters used in 
this paper. Details of the implementation of the lattice Boltzmann method of 
solving the equations can be found in \cite{Pooley08}, and are not 
repeated here.

\subsection{Equations of motion}

We consider a binary fluid with components $A$ and $B$, say, described by 
the free energy functional
\begin{eqnarray}
\Psi = \int_{\Omega} \Big[{c^2\over 3 } n \log n + {\kappa \over 2} 
(\partial_\alpha \phi)^2 - {a\over 2}
\phi^2 + {a\over 4} \phi^4 \Big] + \int_{\partial\Omega} h \cdot \phi_z \, ,
\label{FreeEnergyEq}
\end{eqnarray}
where $n$ is the local total density of the $A$ and $B$ components  
($n=n_A+n_B$), $\phi$ is the order parameter $\phi = n_A - n_B$ and $c$ is the 
lattice velocity $c=\delta x/\delta t$, where $\delta x$ is the lattice spacing 
and $\delta t$ is the simulation time step. The first integral in Eq.\ 
(\ref{FreeEnergyEq}),  taken over the total volume $\Omega$, controls the bulk 
properties of the system.  The terms in $\phi$ give coexistence of phases with 
$\phi=\pm 1$. The energy cost of an interface between the two phases is 
modeled by the derivative term, with $\kappa$ related to the surface tension. 
The term in $n$ controls the compressibility of the fluid.

The integral over the solid-liquid interface $\partial \Omega$ in Eq.\ 
(\ref{FreeEnergyEq}) accounts for the wetting properties of the solid  
surfaces. $h$ is related to the equilibrium contact angle $\theeq$ by  
\cite{C77}
\begin{eqnarray}
h &=& \sqrt{2\kappa a} \cdot \mathrm{sign}\left({\pi\over 2}-\theeq\right)
\sqrt{\cos\left(\alpha\over 3\right) \Big[1- \cos\left(\alpha\over 3\right)\Big]} \, , \nonumber\\
\alpha &=& \mathrm{cos}^{-1} \left( \sin^2 \theeq \right)
\end{eqnarray}
 with $\mathrm{sign}(x)=1$ if $x>0$ and $\mathrm{sign}(x)=-1$ otherwise.

The hydrodynamics of the fluid is described by the  the Navier-Stokes equations
 for the density $\rho$ and the velocity field ${\bf v}$ together with a
 convection-diffusive equation for the binary  order parameter $\phi$ 
\begin{eqnarray}
\partial_t \rho + {\bf \nabla} \cdot (\rho {\bf v}) &=& 0 \, ,
\\
 \partial_t (\rho { v}_\beta) +  \partial_\alpha 
(\rho { v}_\alpha { v}_\beta) &=& -  \partial_\alpha 
[P_{\alpha\beta}  + \eta (\partial_\beta { v}_\alpha + \partial_\alpha
{ v}_\beta)] \, , \label{eq1} 
\\
\partial_t \phi + {\bf \nabla} \cdot (\phi {\bf v}) &=& M   {\bf \nabla}^2 \mu
\, .
\label{eq2}
\end{eqnarray}
In Eq.\ (\ref{eq1}) $\eta$ is the viscosity of the fluid and in Eq.\ 
(\ref{eq2}) $M$ is a mobility coefficient. The pressure tensor $P_{\alpha\beta}$
 and the chemical potential $\mu$ which appear in Eqs.\ (\ref{eq1}) and
(\ref{eq2}), which describe the equilibrium properties of the fluid, follow 
from the free energy (\ref{FreeEnergyEq}) as
\begin{eqnarray}
P_{\alpha\beta} &=& \partial_\alpha \phi {\delta \Psi \over 
\delta (\partial_\beta \phi)} +  \delta_{\alpha\beta} \Big[ 
\phi {\delta \Psi \over \delta ( \phi)} + n {\delta \Psi \over \delta n}
-\Psi \Big]
\nonumber \\
&=& \kappa \partial_\alpha \phi \partial_\beta \phi +\delta_{\alpha\beta}\Big[
{c^2\over 3}n+ {3 a \over 4} \phi^4 -{a\over 2}\phi^2 -{\kappa\over 2} 
(\partial_\tau \phi)^2 - \kappa \phi \partial_\tau \partial_\tau \phi
\Big] \, ,
\\
\mu &=& {\delta \Psi \over \delta \phi} = a\phi^3 - {a\over 2} \phi^2
-\kappa \partial_\tau \partial_\tau \phi \, .
\end{eqnarray}

We have chosen to use a two-component binary fluid as a model system to 
simulate capillary filling. This is because modeling capillary filling 
correctly using a liquid-gas diffuse interface model is computationally 
demanding because of unphysical motion of the 
interface due to evaporation-condensation effects \cite{CondEv}.
However our results are equally applicable to a physical system where a 
liquid displaces a gas as the important physical parameters are the 
viscosities, not the densities, of the fluid components. Therefore we shall 
use the natural terminology `liquid' and `gas' for the displacing and 
displaced fluid from now on.

\subsection{Simulation geometry}

%Fig 1 geometry
\begin{figure}[h]
\includegraphics[angle=0,scale=0.28]{Fig.2a.eps}
\includegraphics[angle=0,scale=0.28]{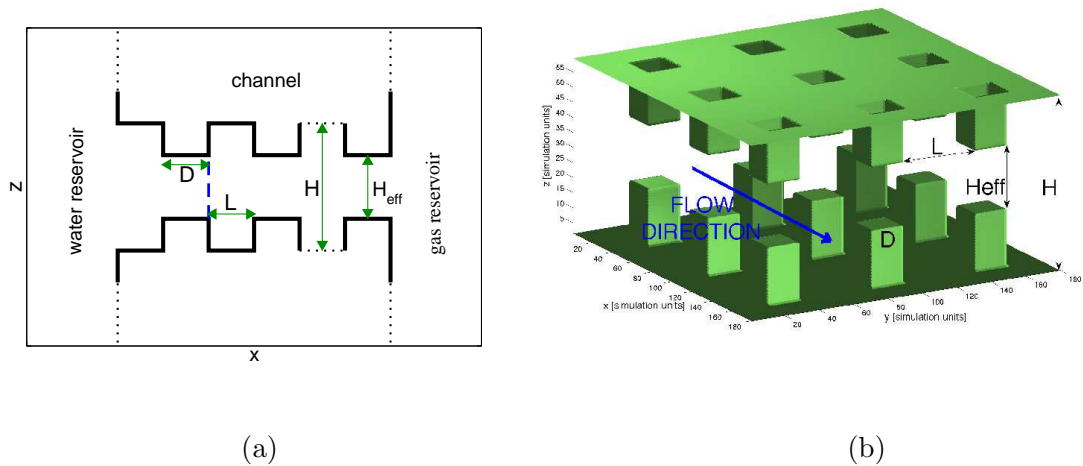}\\ \vspace{0.5cm}
(a) $\qquad\qquad\qquad\qquad\qquad\qquad\qquad\qquad\qquad\quad$ (b)
\caption{ (Color online) Structure and parameters of the channel. The fluid flows from
left to right along the $x$-axis. $H$ is the height of the channel, $H_\mathrm{eff}$ 
the distance between the tops of the posts, $D$ is the dimension of the square posts 
and $L$ is the distance between the posts. Fig.\ \ref{geometry}(a) is a cross section 
in the $x$--$z$ plane, and Fig.\ \ref{geometry}(b) is a three-dimensional view of the 
channel geometry.}\label{geometry}
\end{figure}
Fig.\ \ref{geometry} reports the channel geometry which we use in this paper. 
Two reservoirs, of liquid and gas, which are in contact to equalise the 
pressure, are connected to a channel, running along $x$, with walls decorated 
by equispaced rectangular posts. The relevant geometric parameters are the 
channel height $H$ (measured from wall to wall), the distance between the top 
of two posts on opposing walls $H_\mathrm{eff}$, the distance between two 
obstacles across the channel $L$, and the cross section of the 
posts which we choose to be square with side of length $D$.

\subsection{Simulation parameters}

All the quantities reported in this paper are expressed in units of 
$\delta x$, the lattice spacing, and $\delta t$, the time step and
hence $c=1$. 
We simulate channels with one, two or three rows of posts corresponding
to lengths (in the $x$ direction, see Fig.\ \ref{geometry}) $L_x$ from $80$
to $200$ lattice spacing. In the $y$ direction we employ periodic boundary 
condition and sizes $L_y$ from $60$ to $170$, while $H=L_z$ spans from $30 $ 
to $100$. For the reservoirs we use the same 
$L_{y,\mathrm{res}}=L_y$ while in the 
$z$ direction we use periodic boundary conditions with 
$L_{z,\mathrm{res}}$  two or three times $L_z$. The $x$-length of the reservoir
$L_{x,\mathrm{res}}$ is approximately half the channel length $L_x$.
For the free energy (\ref{FreeEnergyEq}), the bulk phases ($\phi=1$ 
and $\phi=-1$) are interpolated by an interface with a profile which is 
well approximated by $\phi = \tanh(x/\sqrt{2}\xi)$ (with $\xi=\sqrt{k/a}$), 
and with a surface tension equal to $\gamma=\sqrt{8 k a/9}$. Here we use 
$a=0.04$ and $k=0.02$. These values give an interface width of order four 
lattice Boltzmann nodes which is much smaller than the typical size of the 
posts used.
The viscosity  $\eta$ and the mobility coefficient $M$ appearing in the 
hydrodynamic equations (\ref{eq1}, \ref{eq2}) are related to the relaxation 
time in the lattice Boltzmann algorithm \cite{Succi}. 
We use $M=0.5$.  For the  gas 
viscosity $\eta_\mathrm{gas}=0.033$, while for the liquid (unless specified) 
$\eta_\mathrm{liq}=0.83 $. Both the gas and liquid densities are set to one.

\section{Results}\label{secRes}

%Fig 3 positon of interface as function of time
\begin{figure}[h]
\includegraphics[angle=-90,scale=0.29]{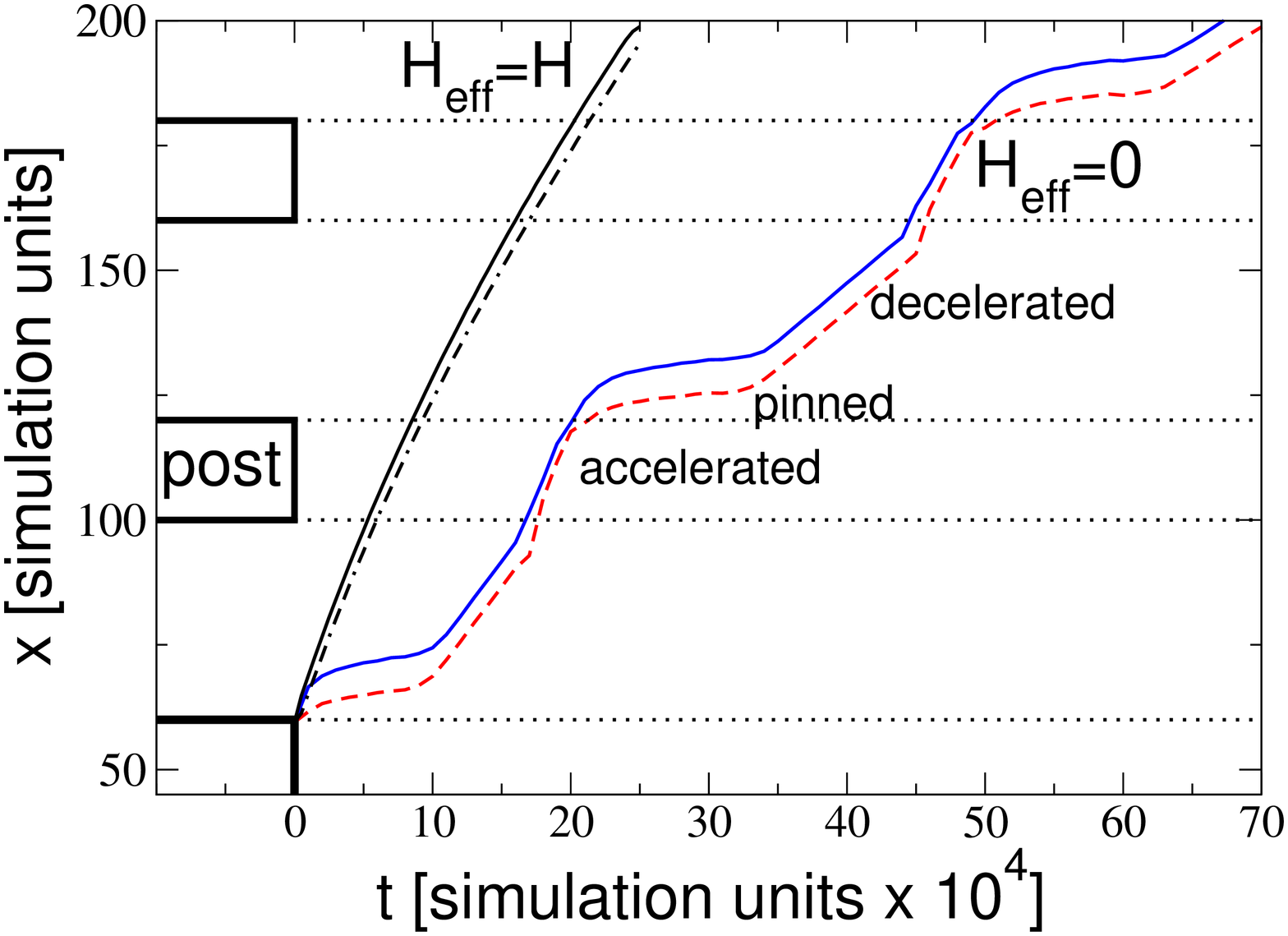}
\includegraphics[angle=-90,scale=0.29]{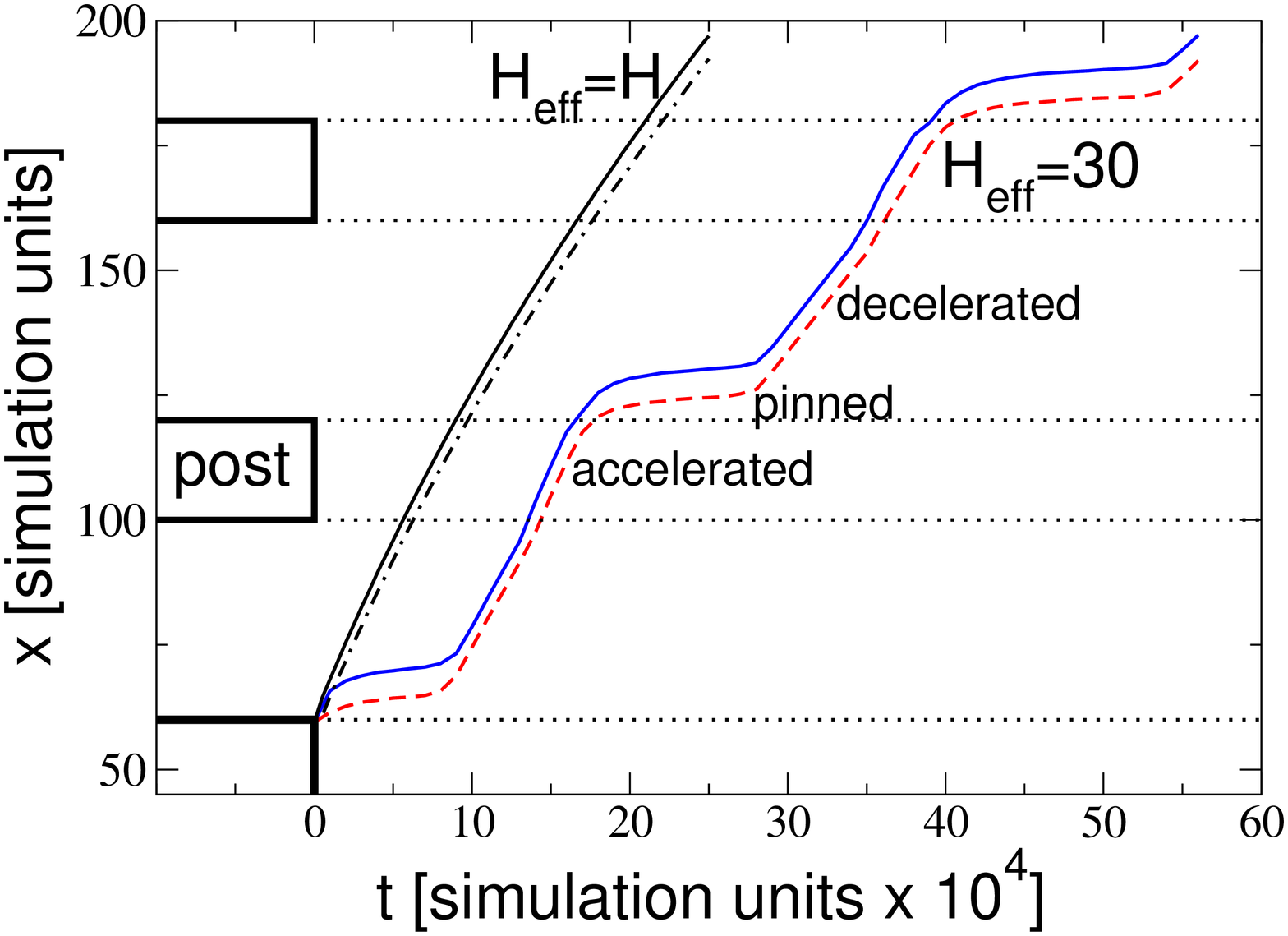}\\
(a) $\qquad\qquad\qquad\qquad\qquad\qquad\qquad\qquad\qquad\quad$ (b)
\caption{(Color online) Position $x$  of
the advancing front 
for (a) $D=20$, $L=40$, $H_\mathrm{eff}=0$, $H=40$, $\theeq=55^\circ$; (b)
  $D=20$, $L=40$, $H_\mathrm{eff}=30$, $H=50$, $\theeq=60^\circ$ 
as a function of time $t$. The position of the advancing front near the wall  
(full line) and in the middle of the channel 
(broken line) are compared. We compare
 the filling rate for the same channel without posts 
($H_\mathrm{eff}=H$).}\label{threePosts}
\end{figure}	
In a typical simulation we start with a configuration in which the liquid-gas 
interface has advanced to the end of the first row of posts as indicated by the 
broken line in Fig.\ \ref{geometry}. If the contact angle is not too large, the 
interface moves through the channel,
 driven by the capillary force. If the filling 
fluid  overcomes the first row of obstacles, we have verified it depins from 
the second row too. We have also used starting configurations with empty channels,
finding no differences in the filling/pinned phase diagram.

Fig.\ \ref{threePosts} shows the position of the advancing front (in the middle 
and near the walls of the channel) as a function of time for two typical 
geometries and for contact angles which do not pin. It is immediately apparent 
that the flow profile is very different to that of a smooth channel 
$H_\mathrm{eff}=H$. Three regimes are present. When the advancing fluid
reaches the beginning of the obstacles (dotted lines in Fig.\ \ref{threePosts})
it accelerates because of the increase in the 
capillary force due to more wettable 
surface provided by the obstacles. When the front reaches the end of the obstacles 
it remains pinned for a certain time during which it is almost at rest.
When finally it depins it restarts filling the channel with a Lucas Washburn-like 
law (\ref{WashburnEq}), but with a reduced velocity (the ``decelerated''
regime compared to the smooth channel in Fig.\ \ref{threePosts}). This happens because  the drag force
is now larger due to the presence of the obstacles within the
displacing viscous fluid. We observe that the ``accelerated'' 
and ``decelerated'' regimes are more obvious for  $H_\mathrm{eff}=0$,
because the post
surface is bigger than for $H_\mathrm{eff}=30$.

For the parameters considered in Fig.\ \ref{threePosts} the front is finally 
able 
to depin from each of the obstacles to move down the channel, but for lower 
contact angles it remains pinned. We next look in more detail at the pathways 
for depinning.

\subsection{$H_\mathrm{eff}=0$ }\label{secHeff0}

\begin{figure}[h]
\includegraphics[angle=-90,scale=0.4]{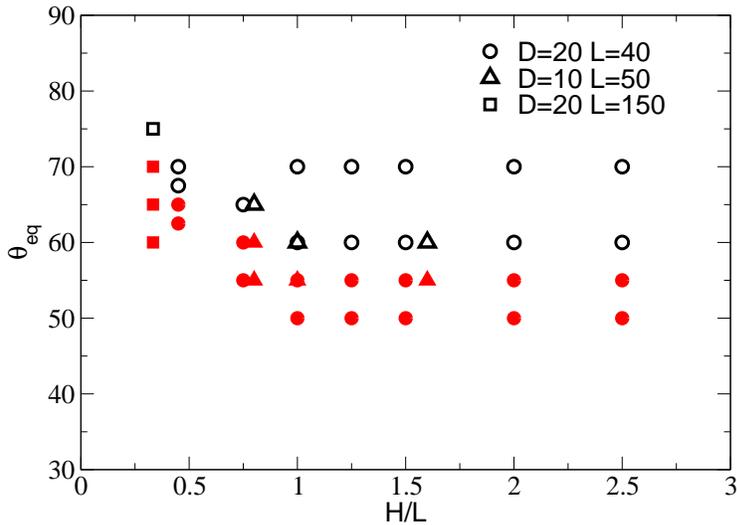}
\caption{(Color online) Filling/pinning transition for the 
$H_\mathrm{eff}=0$  geometry (posts that span the channel, 
see Fig.\ \ref{geometry}). 
Open symbols are simulations in which
the front is pinned while full symbols correspond to simulations 
in which the channel is filled.  For $H/L\gtrsim 1$ the filling-pinning
 transition happens for a threshold angle $\sim 55^\circ$ while for 
$H/L\lesssim 1$ the transition is possible for a range of equilibrium contact 
angles which increases with decreasing $H/L$.
}\label{fullobs}
\end{figure}
We first consider geometries in which the posts span the channel 
($H_\mathrm{eff}=0$ in Fig.\ \ref{geometry}).  Fig.\ \ref{fullobs}
summarises results which distinguish the cases where the interface is 
pinned on the posts from those where it can advance along the channel, for a 
range of geometric parameters  ($H$, $L$ and $D$, defined in  Fig.\ 
\ref{geometry}) and equilibrium contact angles $\theeq$.

\begin{figure}[h]
\includegraphics[angle=0,scale=0.36]{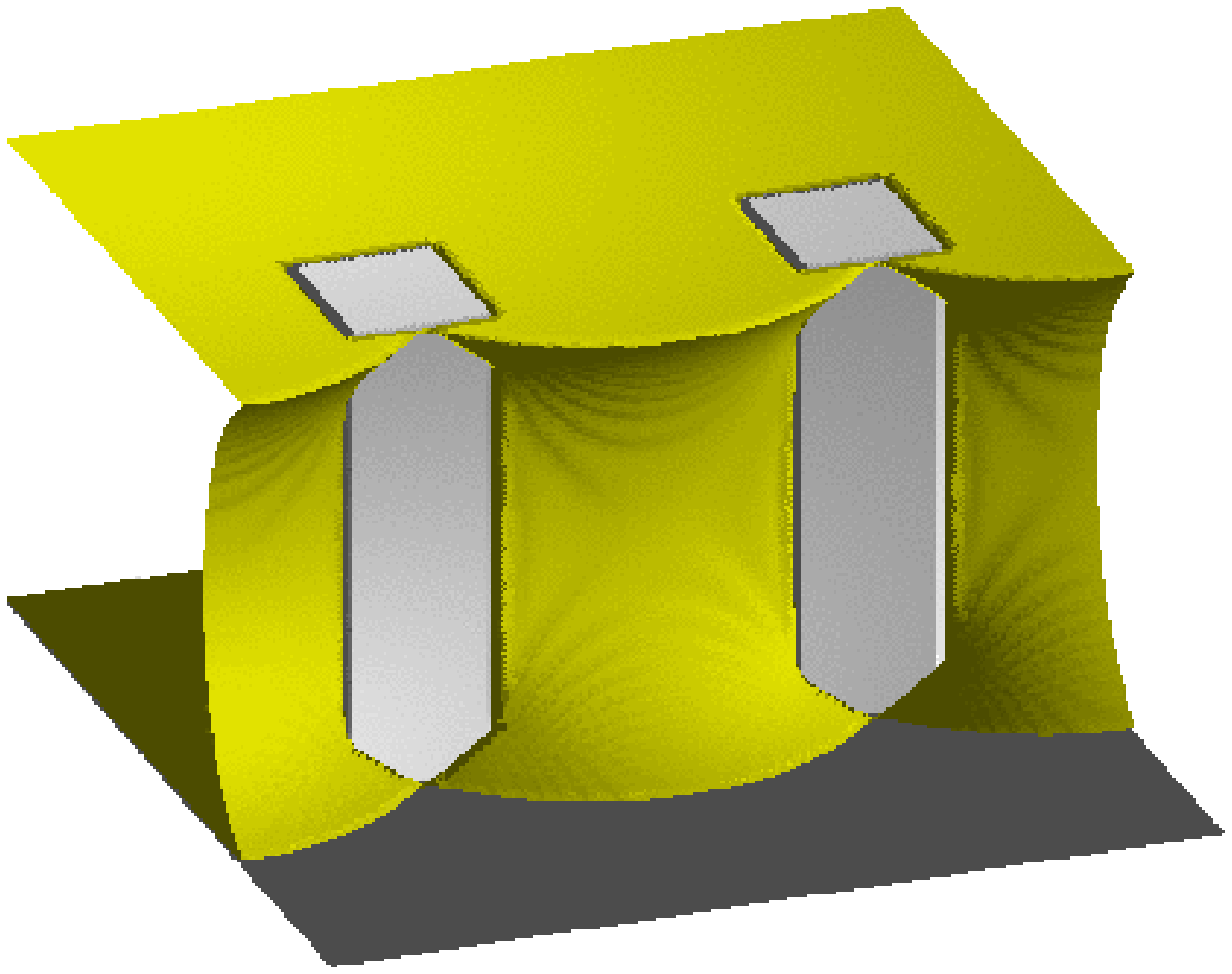}
\includegraphics[angle=0,scale=0.25]{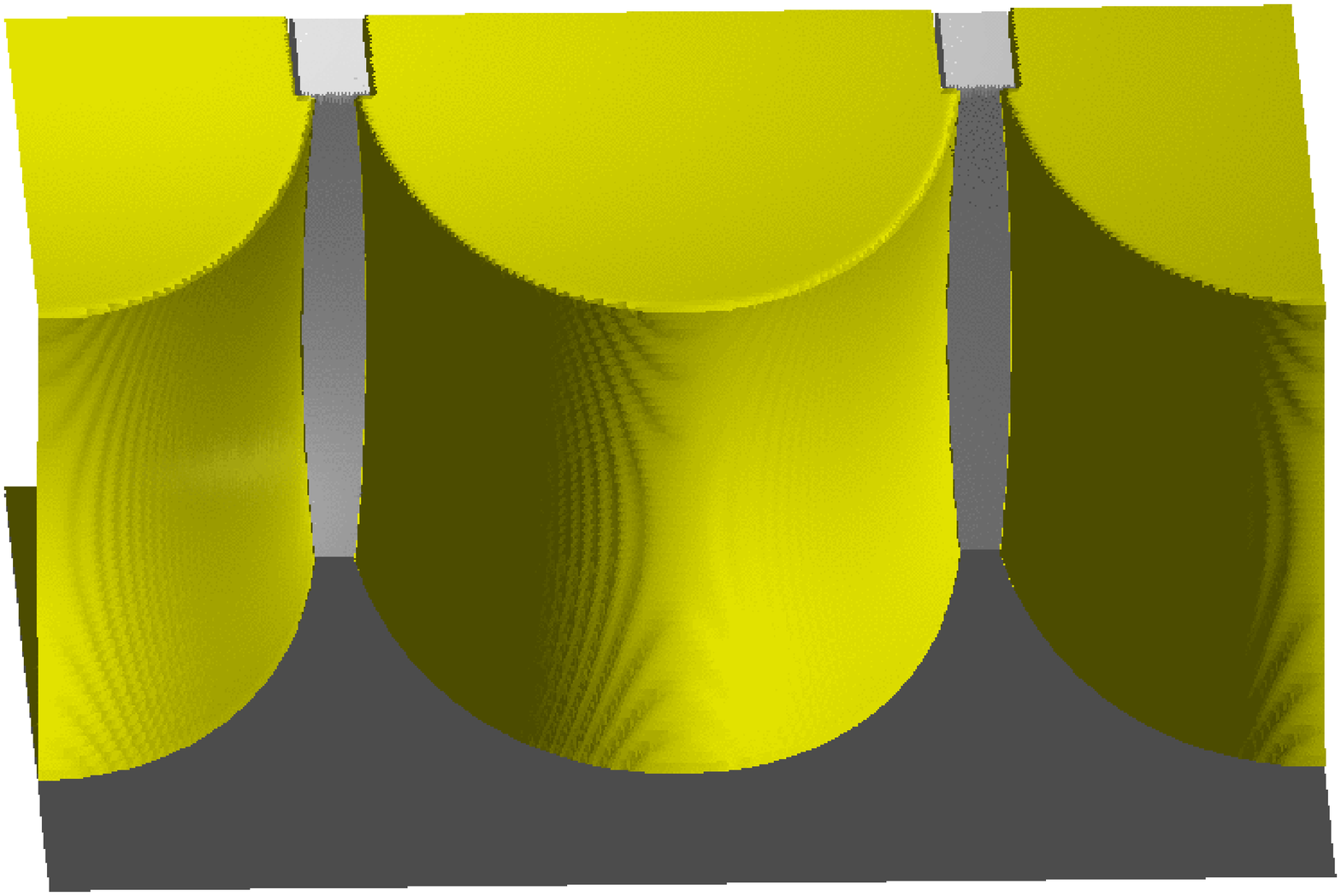}
\includegraphics[angle=0,scale=0.36]{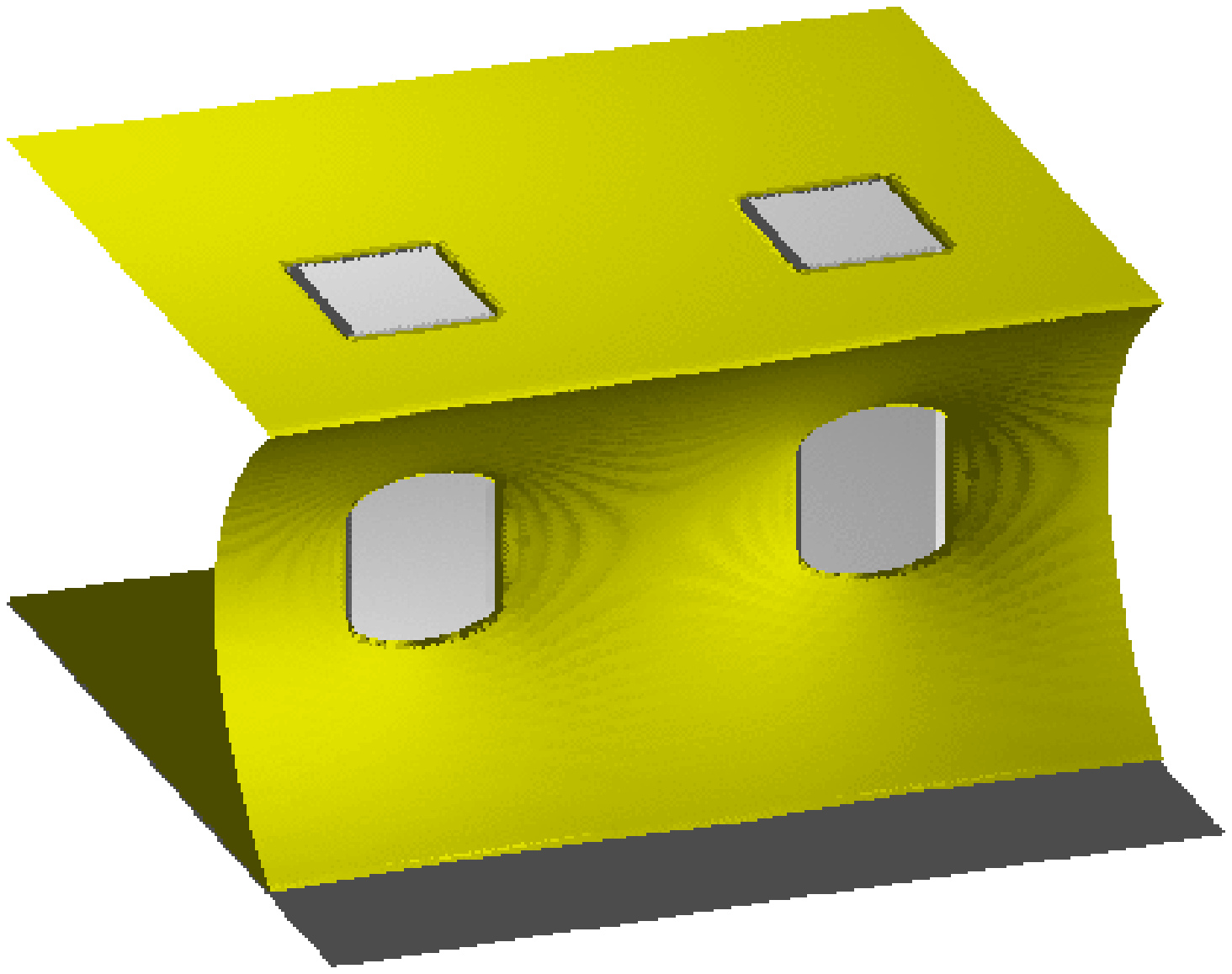}
\includegraphics[angle=0,scale=0.25]{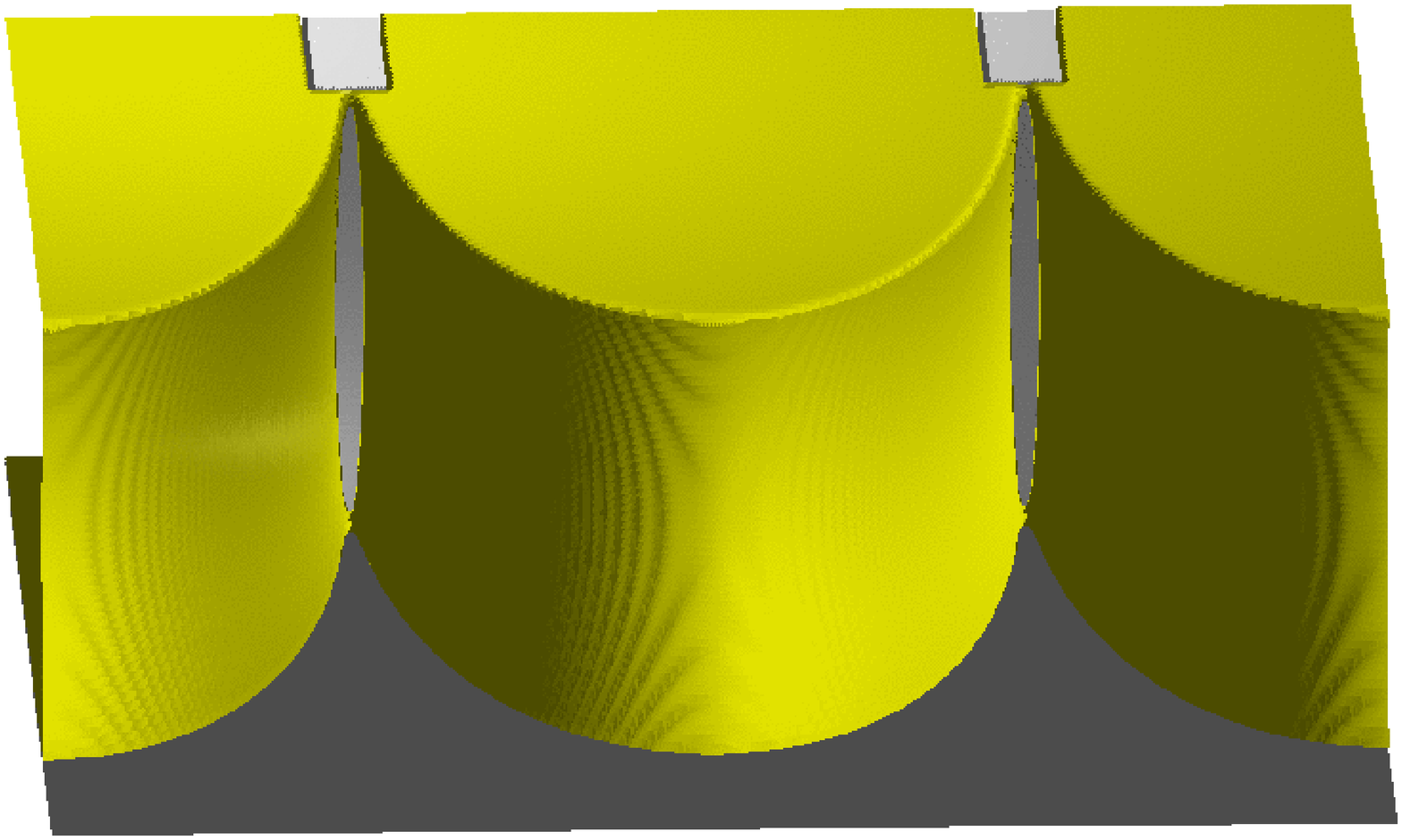}
\includegraphics[angle=0,scale=0.36]{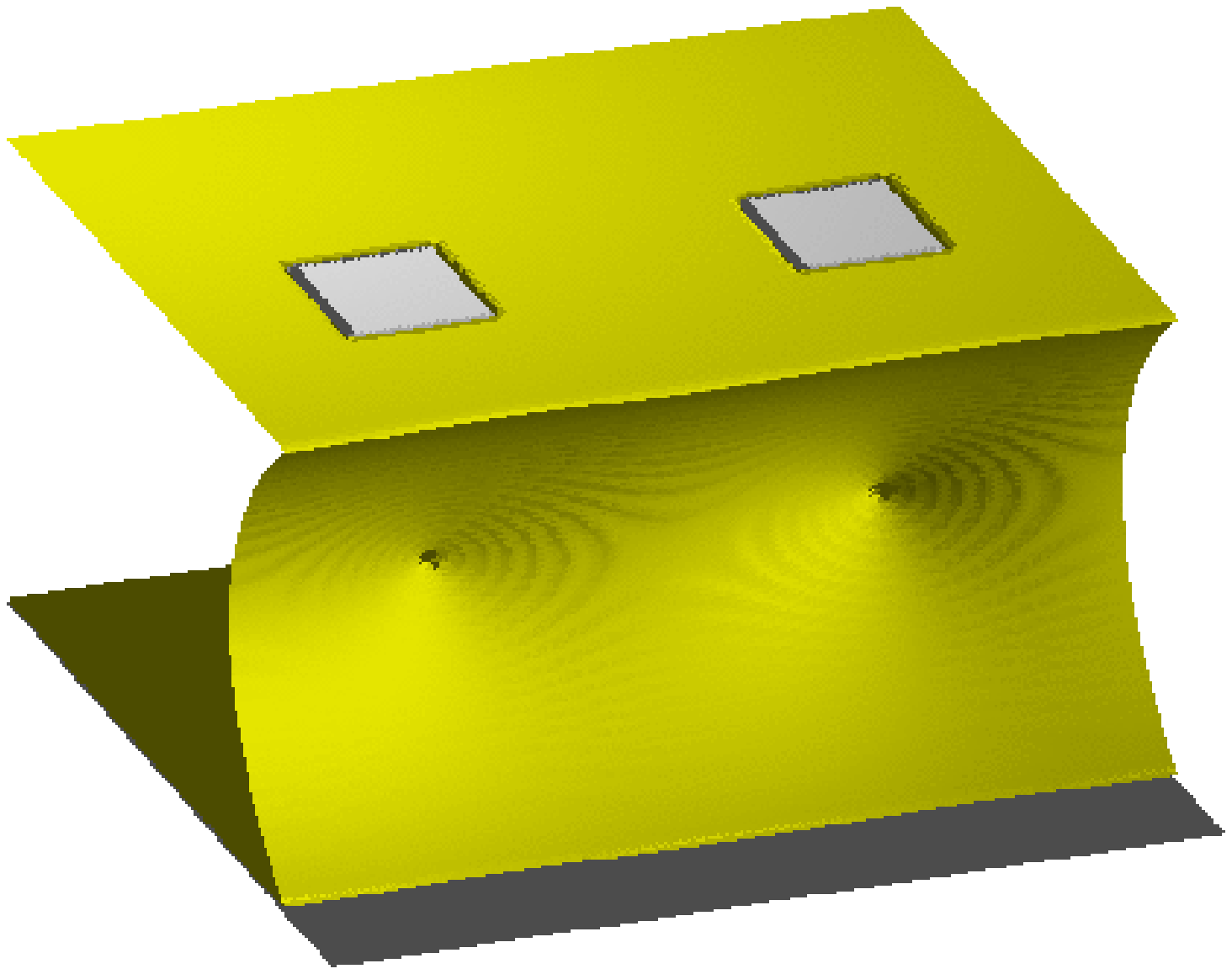} 
\includegraphics[angle=0,scale=0.25]{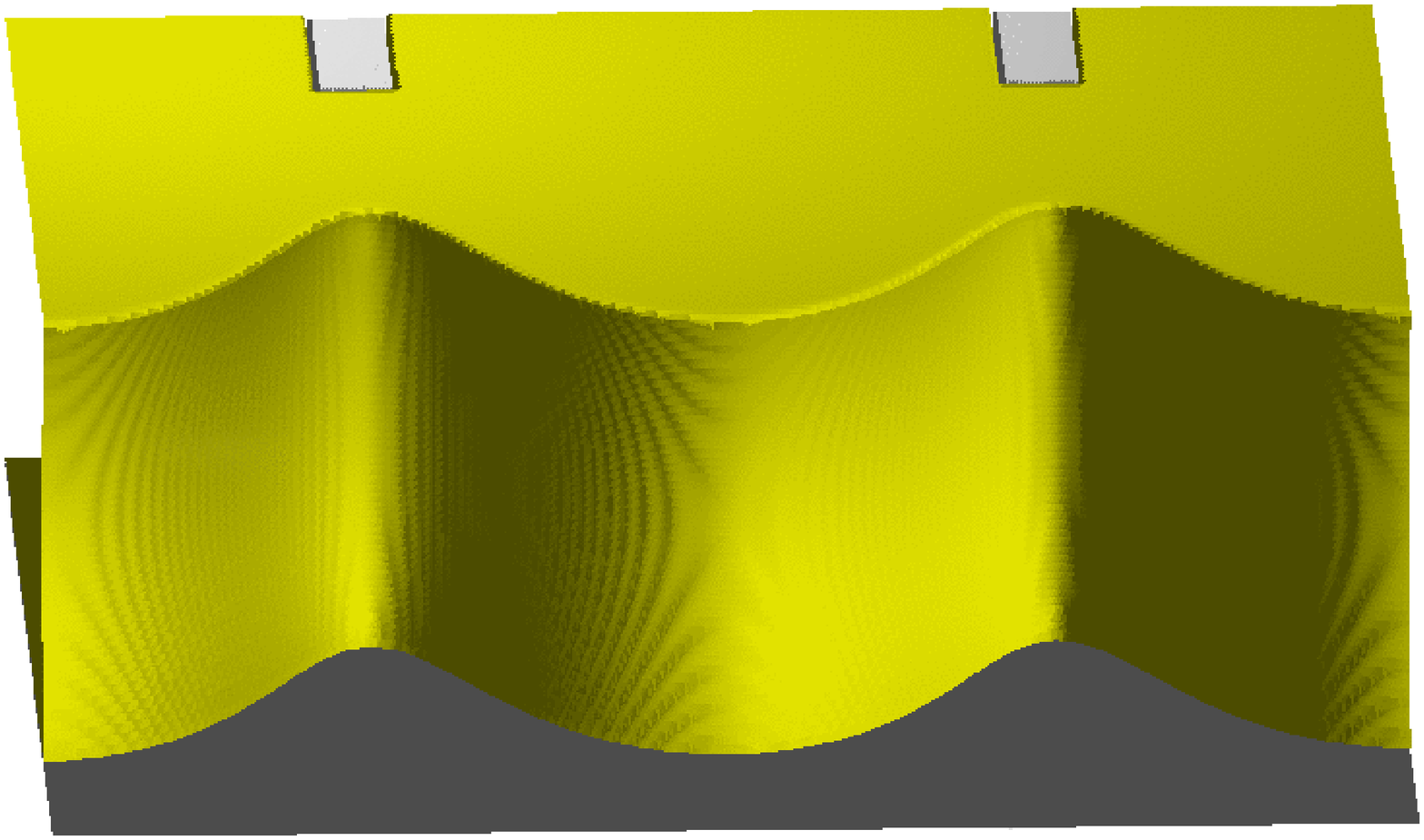}     
\includegraphics[angle=270,scale=0.23]{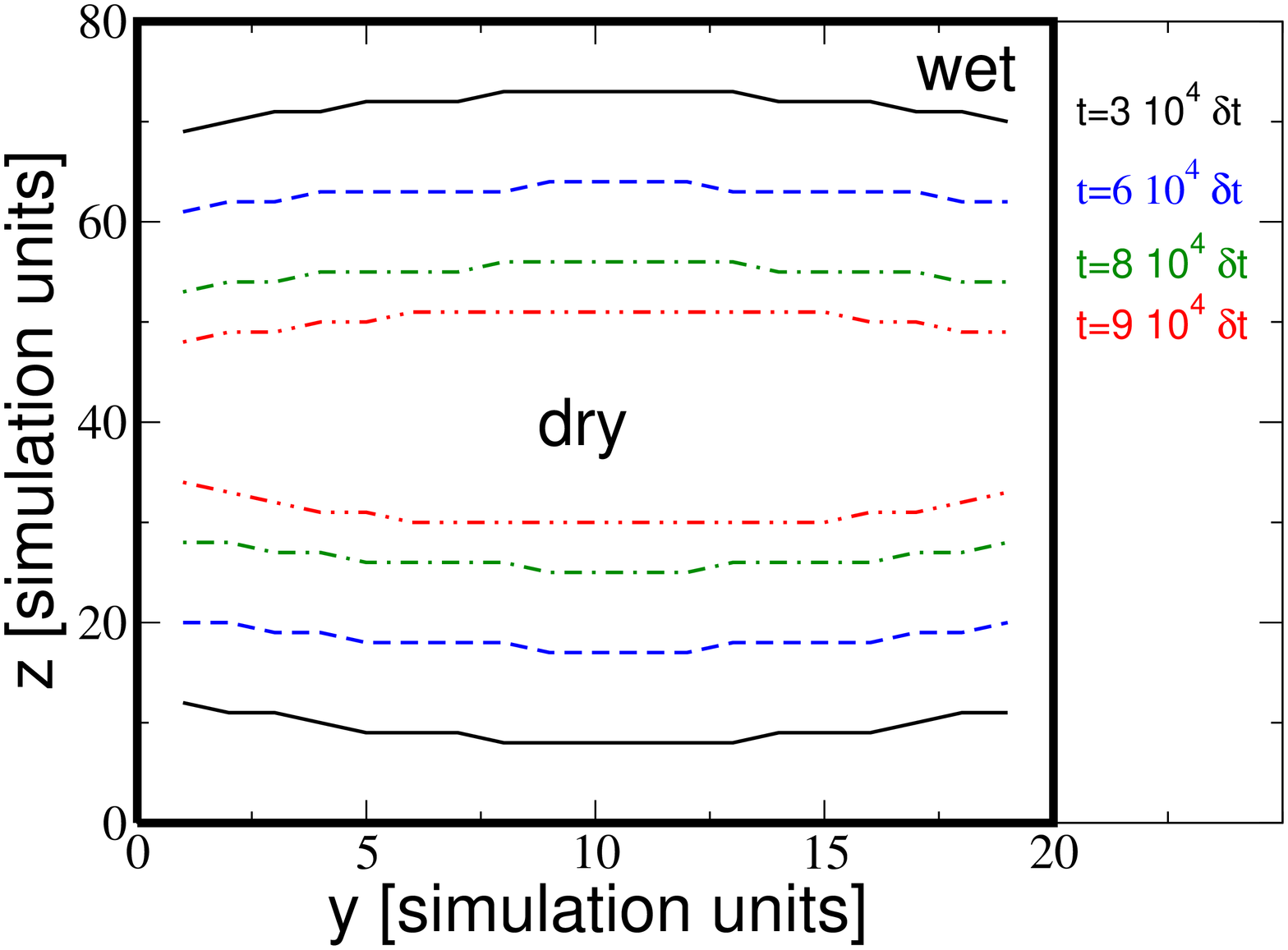}
\includegraphics[angle=270,scale=0.23]{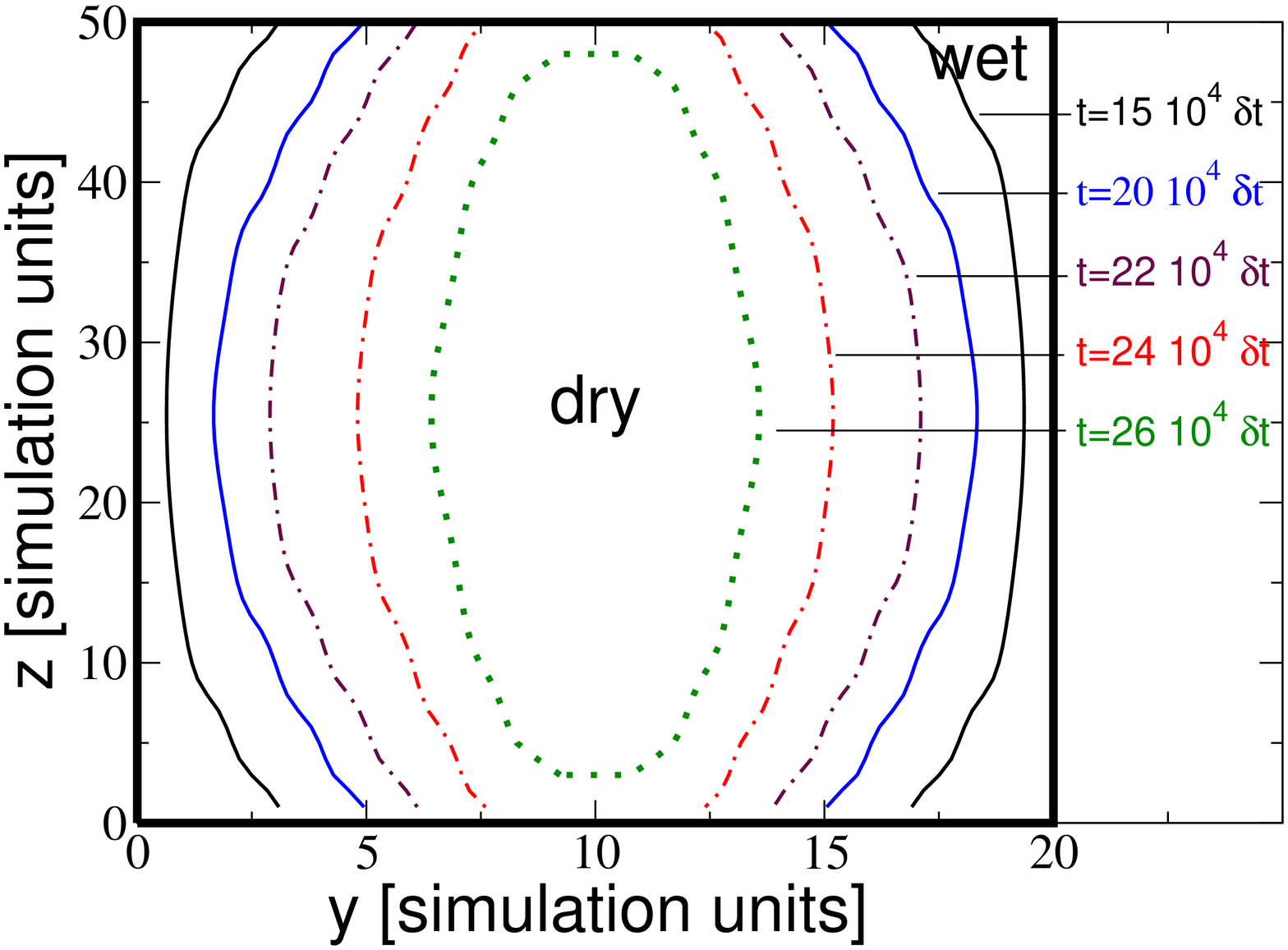}\\
(a) $\qquad\qquad\qquad\qquad\qquad\qquad\qquad\qquad\qquad\quad$ (b)
\caption{
(Color online) Depinning path ways for $H_\mathrm{eff}=0$: (a)
$H/L>1$ ($D=20$, $L=40$,  $H=80$, $\theeq=50^\circ$)  
and  (b) $H/L<1$ ($D=20$, $L=150$,  $H=50$, $\theeq=70^\circ$). 
 The first three rows are three dimensional views of the 
advancing front (aspect ratios not to scale) and the fourth row shows
the position of the advancing front on the face of the posts as a function
of time during depinning. Note that for $H/L\gtrsim 1$ (a) the fluid wets 
the posts from  the walls of the channel whereas for $H/L\lesssim 1$ (b) 
the fluid  
advances from the sides of posts.
}\label{dep0}
\end{figure}

At high $\theeq$ the front remains pinned. However, at lower $\theeq$ the 
meniscus can overcome Gibb's pinning and the channel fills. This is due to 
the presence of walls bounding the channel. By advancing along the walls 
the meniscus is able to reach the angle it needs to 
move across the face of the posts. This is illustrated in Fig.~\ref{dep0}
for different aspect ratios of the channel.

Two different regimes are apparent in Fig.~\ref{geometry}. For $H/L \gtrsim 1$
the boundary between contact angles that allow filling and those that do not 
is independent of $H/L$ occurring at an angle, that we shall denote
 $\thetr$, $\sim 55^\circ$. For $H/L 
\lesssim 1$, however, the transition angle is not constant but increases with 
decreasing $H/L$.

Fig.~\ref{dep0} compares the way in which the front depins from the posts in 
each of the two regimes.  For $H/L\gtrsim 1$ (Fig.~\ref{dep0} column (a)) the walls 
act independently. The menisci from two neighboring gaps first meet at the 
walls and then the advancing front covers the posts, moving from the walls 
towards the centre of the channel. Hence the contact angle below which 
depinning proceeds, $\thetr$, is independent of $H/L$.

\begin{figure}[h]
\includegraphics[angle=-90,scale=0.5]{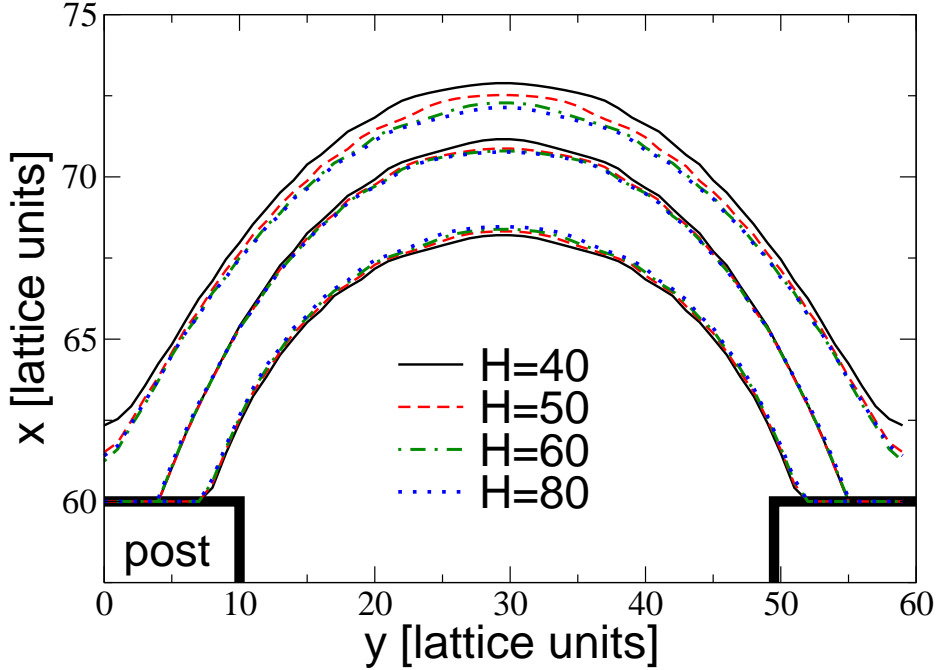}
\caption{(Color online) Position of the depinning front near the
walls (at $z=2$), as a function of time (for $t=1,2$ and $3\cdot 10^4$ 
 lattice Boltzmann time steps) for $\theeq=50^\circ$ and several
$H$ so that $H/L\gtrsim 1$ with $L=40$. 
 }\label{adv_z1}
\end{figure}
Further evidence is provided for this by Fig.~\ref{adv_z1} which shows 
how the meniscus advances along the wall for different values of $H$ and at 
fixed $L$. The profiles are nearly independent of $H$ as long as the front remains 
pinned. Without obstacles, Lucas-Washburn's law  predicts a velocity proportional to 
$H$ (Eq.\ (\ref{WashburnEq})), so that the profiles in Fig.\ \ref{adv_z1} would 
be well separated in a smooth channel.

We stress the importance of the walls even in the limit $H\to \infty$. 
Even
for a very high channel depinning  will occur at $\thetr$, 
and will proceed as in Fig.\ 
\ref{dep0} column (a) (except that it will take  more time to wet the posts). 
However, without any walls the advancing fronts will remain pinned  at the
obstacles and flat as in Fig.\ \ref{GibbsFig}(b).

For $H/L\lesssim 1$, however, 
depinning occurs for contact angles greater than $\thetr$.
This is because  the walls are sufficiently close  that the 
interface moves in a concerted way across the channel. 
Therefore, once the menisci have advanced along the 
surfaces sufficiently far for depinning to occur, the interface depins along 
all of a post at the same time, and the posts are wet from the sides. As $H/L$ 
decreases the two surfaces more easily deform the interface and hence depinning
 can take place at a higher contact angle. This agrees with analytic results 
showing that, in the $H/L \rightarrow 0$  limit,  the interface depins for 
all hydrophilic contact angles \cite{HS-ST}, 
and can be understood
 using a free energy argument. The free energy gain in deforming the advancing 
interface  scales like the area of the gas-liquid interfaces, 
and goes to zero as $H$ for 
$H\to 0$. However  the loss in free energy due to  the wetting of
the walls remains constant for $H\to 0$. As a consequence the interface 
will advance for any  $\theeq<90^\circ$ for sufficiently small $H$.

We expect 
the phase diagram to depend only on  $H/L$ and to be independent of $D$ as, 
once the front has started to depin, the interface will continue moving across 
the post until it covers it. 
Indeed, using $L=40$ we repeated simulations for $D=30$ ($H=60$, 80)
and $D=40$ ($H=60$, 80) finding, again, a depinning transition compatible 
with $\thetr$.
On the other hand we observed a weak dependence of $\thetr$ on $L$. We
repeated simulations for $L=30$, $D=30$ and $H/L=20$, 30, 50, 80, 100 and
120. $\thetr$ was again independent of $H/L$ for $H/L\gtrsim 1$, but
placed between $50^\circ$ and $55^\circ$, while for $L=40$ (Fig.\ \ref{dep0})
$\theeq = 55^\circ$ fills the channel.
In Sec.\ \ref{secIn} we show the dependence of $\thetr$ on $L$ is 
related to inertial effects.

\subsection{$H_\mathrm{eff}>0$ }\label{secHeff}

In this section we describe the behaviour for the more general case 
$H_\mathrm{eff}>0$ when the posts do not reach all the way across the 
channel. We will consider the $H/L \gtrsim 1$ geometry where the interface 
depins from posts with $H_\mathrm{eff}=0$ by a transition at $\thetr$, driven 
by depinning initiated at the walls. For $H_\mathrm{eff}>0$ filling the channel 
should be easier.

\begin{figure}[h]
\includegraphics[angle=-90,scale=0.5]{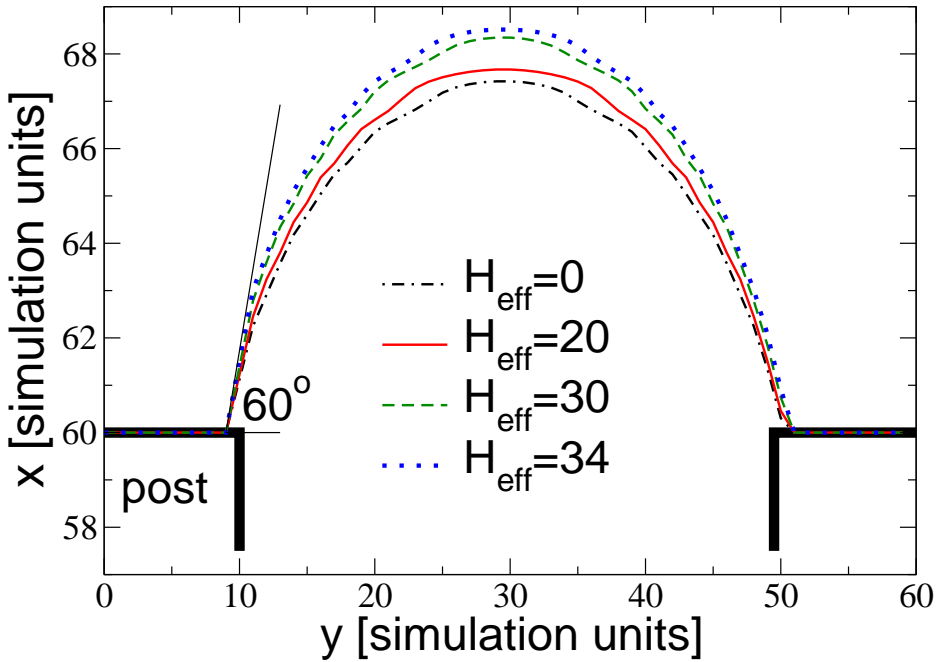}
\caption{(Color online) Position of the pinned meniscus near the walls 
($z=2$) for $H=60$, $D=20$, $L=40$, $\theeq=60^\circ$
and four values of $H_\mathrm{eff}$. Increasing $H_\mathrm{eff}$
the contact angle at the edge of the post is increased until the
depinning threshold is reached (for $H_\mathrm{eff}=40$).
 }\label{AngularRise}
\end{figure}
Fig.\ \ref{AngularRise} shows pinned configurations of the meniscus 
near the channel wall for different values of $H_\mathrm{eff}$, keeping 
the other geometric parameters fixed, and for $\theeq=60^\circ$. For 
$H_\mathrm{eff}=0$  this corresponds to a pinned configuration. The effect of 
increasing $H_\mathrm{eff}$ is to increase the contact angle at the edge of 
the post. When this angle exceeds the equilibrium contact angle 
($\theeq=60^\circ$ in Fig.\ 
\ref{AngularRise}) then, as predicted by the Gibbs' criterion, 
the front depins. 
For the parameters of Fig.\ \ref{AngularRise} this happens for 
$H_\mathrm{eff}=40$ (not reported in the figure).

For $H_\mathrm{eff} \lesssim H/2$, the angle the meniscus makes with the
edge of the post does not 
change significantly. For this reason, at low values of $H_\mathrm{eff}$
the threshold equilibrium contact angle is almost the same as in the 
$H_\mathrm{eff}=0$ case (i.e.\ $\approx\thetr$). However the meniscus advances
significantly from its position at $H_\mathrm{eff}=0$ when 
$H_\mathrm{eff}$ becomes comparable to $H$. This can be explained by
considering  the shape of the pinned interface. If $H_\mathrm{eff}=0$ 
and $H/L\gtrsim 1$, the pinned front is almost flat in the middle of the channel 
with significant deviations only near the walls. 
If $H_\mathrm{eff}$ is small enough that the gap does not overlap with the wall
regions, then it will not perturb the interface compared to the $H_\mathrm{eff}
=0$ configuration.
 On the other hand for large $H_\mathrm{eff}$, the end of the posts will lie
within the wall region and there will be significant deformation of the pinned 
front, compared to the $H_\mathrm{eff}=0$ case, and a threshold equilibrium 
contact angle different to $\thetr$.

For several channels ($D=20$, $L=40$, $H=40$, 50, 60, 70, 80) we have verified 
that at low $H_\mathrm{eff}$ the front remains pinned at $\theeq=60^\circ >
\thetr$, as in the $H_\mathrm{eff}=0$ case, while at high enough 
$H_\mathrm{eff}$, $\theeq=60^\circ$ can fill the channel.
 We were never able to observe filling for $\theeq=65^\circ$, but  
we do not exclude that at high enough $H_\mathrm{eff}$ (which would require
a larger simulation box to properly resolve the small height of the posts) 
this can happen.

\begin{figure}[h]
\includegraphics[angle=-90,scale=0.35]{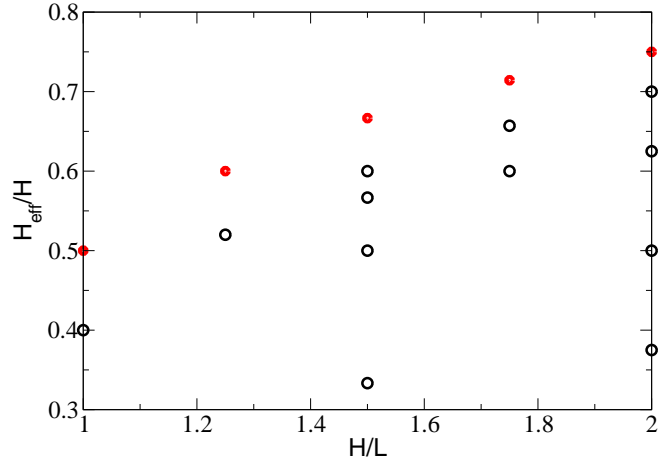}
\caption{(Color online) 
Full  symbols denote values of $H_\mathrm{eff}$  
(with $D=20$, $L=40$ and $H=40,50,60,70,80$) for which the channel will
fill at an equilibrium contact angle $\theeq=60^\circ$.  
Empty circles, correspond to interfaces that remain pinned at $\theeq=60^\circ$
as in the case $H_\mathrm{eff}=0$.}\label{HEff} 
\end{figure}
Results summarising the threshold in $H_\mathrm{eff}$ above which 
$\theeq=60^\circ$
fills are reported in Fig.~\ref{HEff}. Having fixed $D$ and $L$, we use 
$H_\mathrm{eff}/H$ and $H/L$ as control parameters. At low value of $H/L$ the 
filling for $\theeq=60^\circ$ occurs for values of $H_\mathrm{eff}/H$ greater 
than 0.5 (see full circles). Increasing $H/L$ higher values of  
$H_\mathrm{eff}/H$ are needed to achieve filling at $60^\circ$ because
the length of flat interface in the middle of the channel 
increases with increasing $H$.

\begin{figure}[h]
\includegraphics[angle=0,scale=0.4]{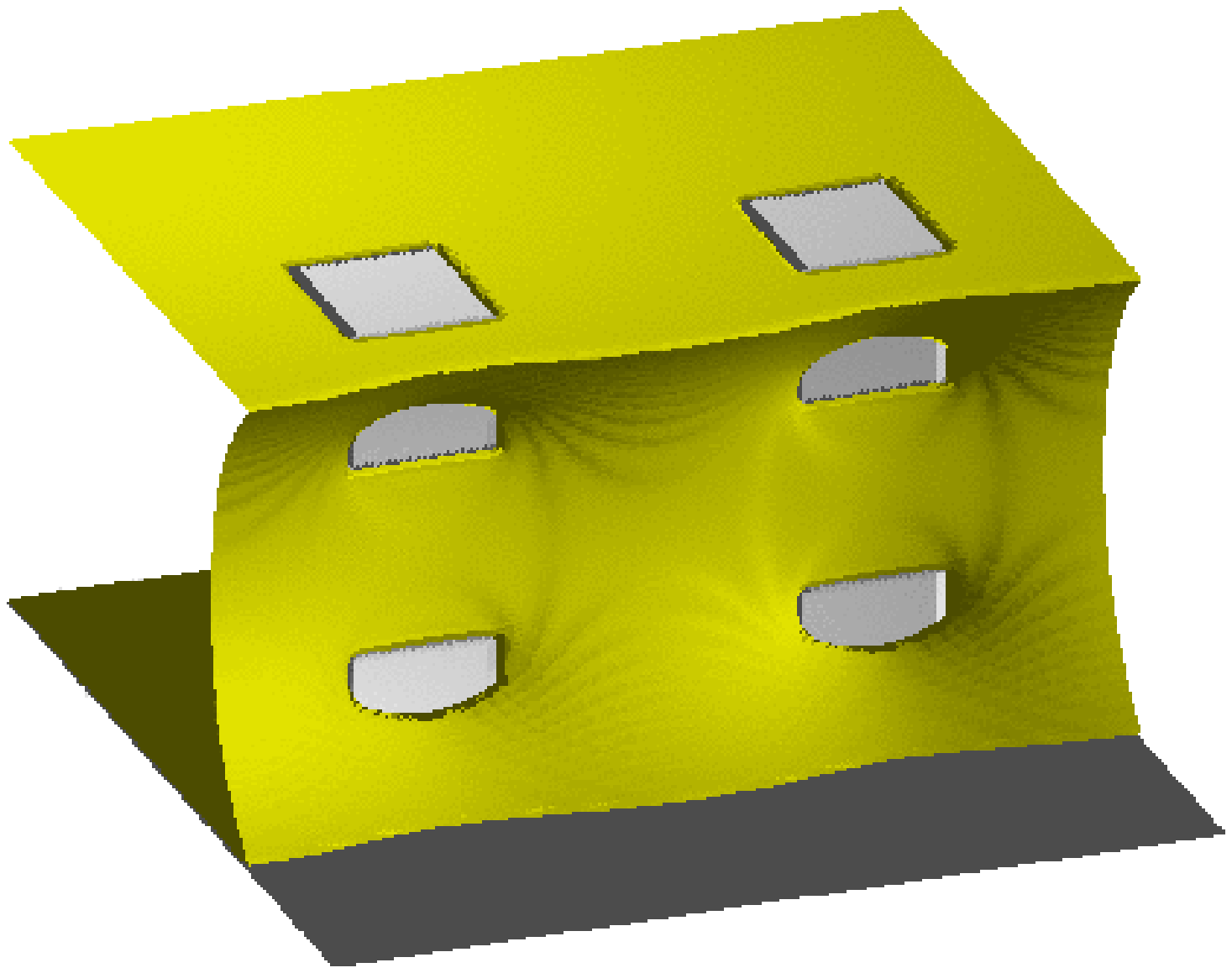}
\includegraphics[angle=0,scale=0.4]{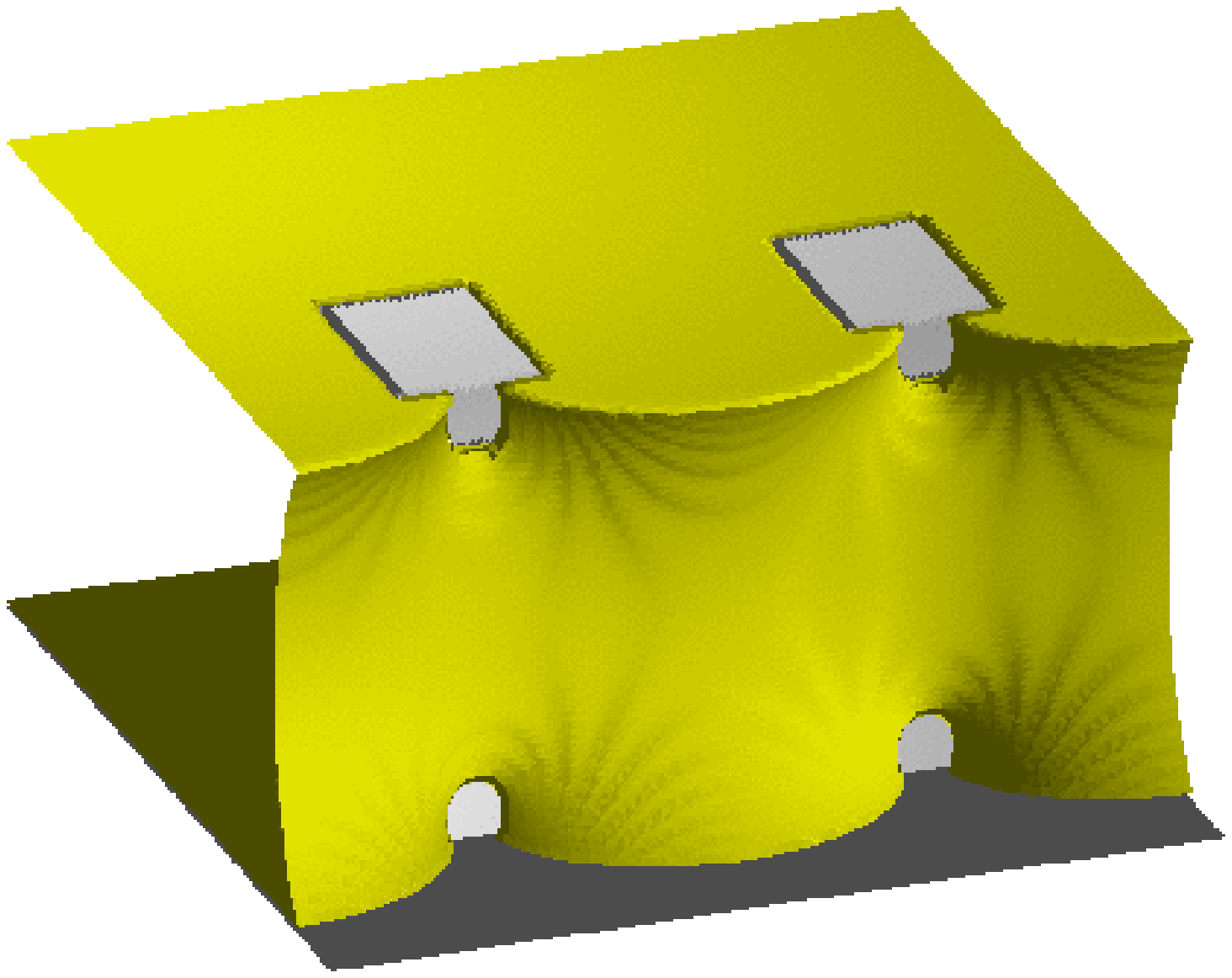}\\
\includegraphics[angle=-90,scale=0.27]{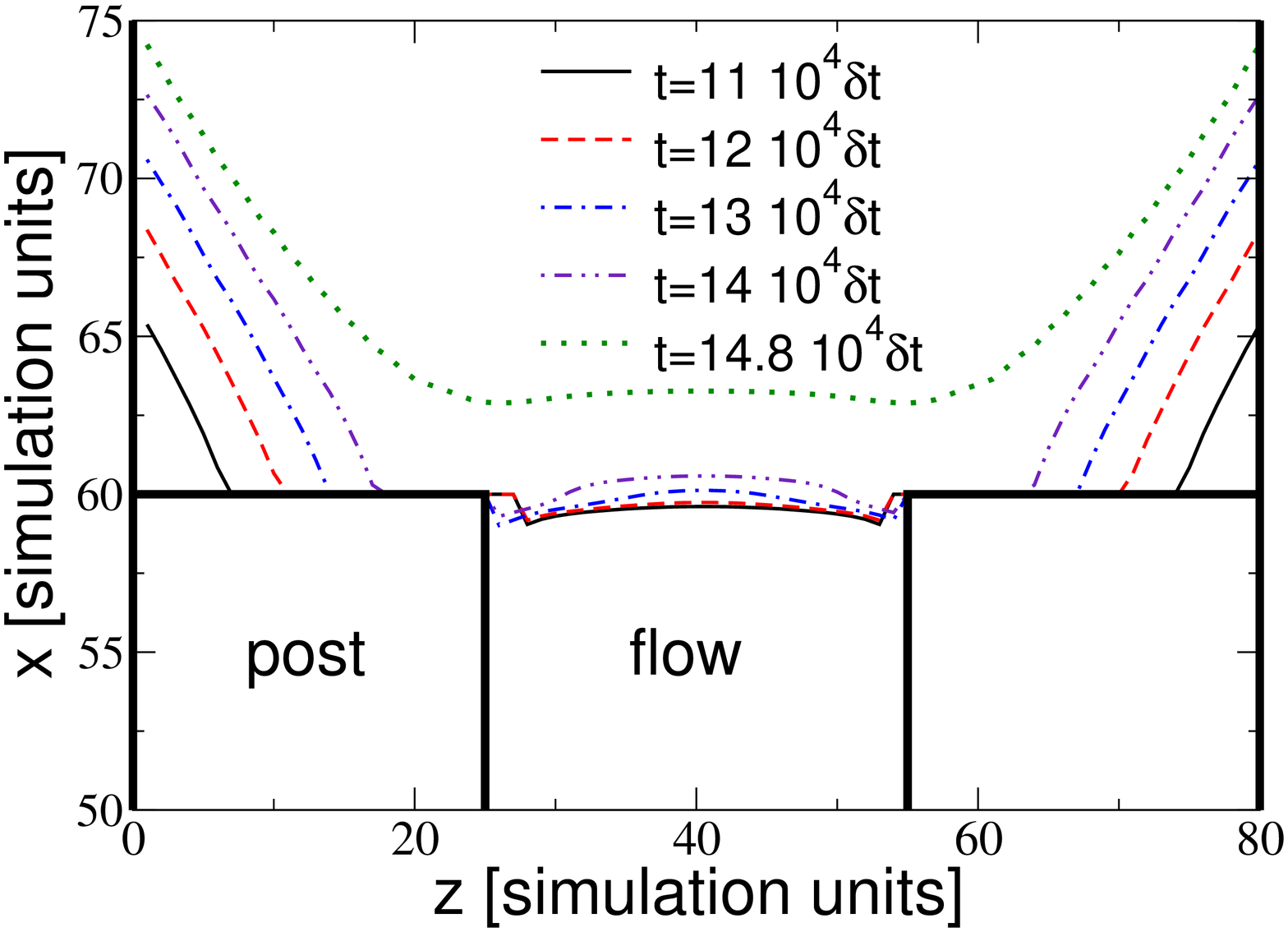}
\includegraphics[angle=-90,scale=0.27]{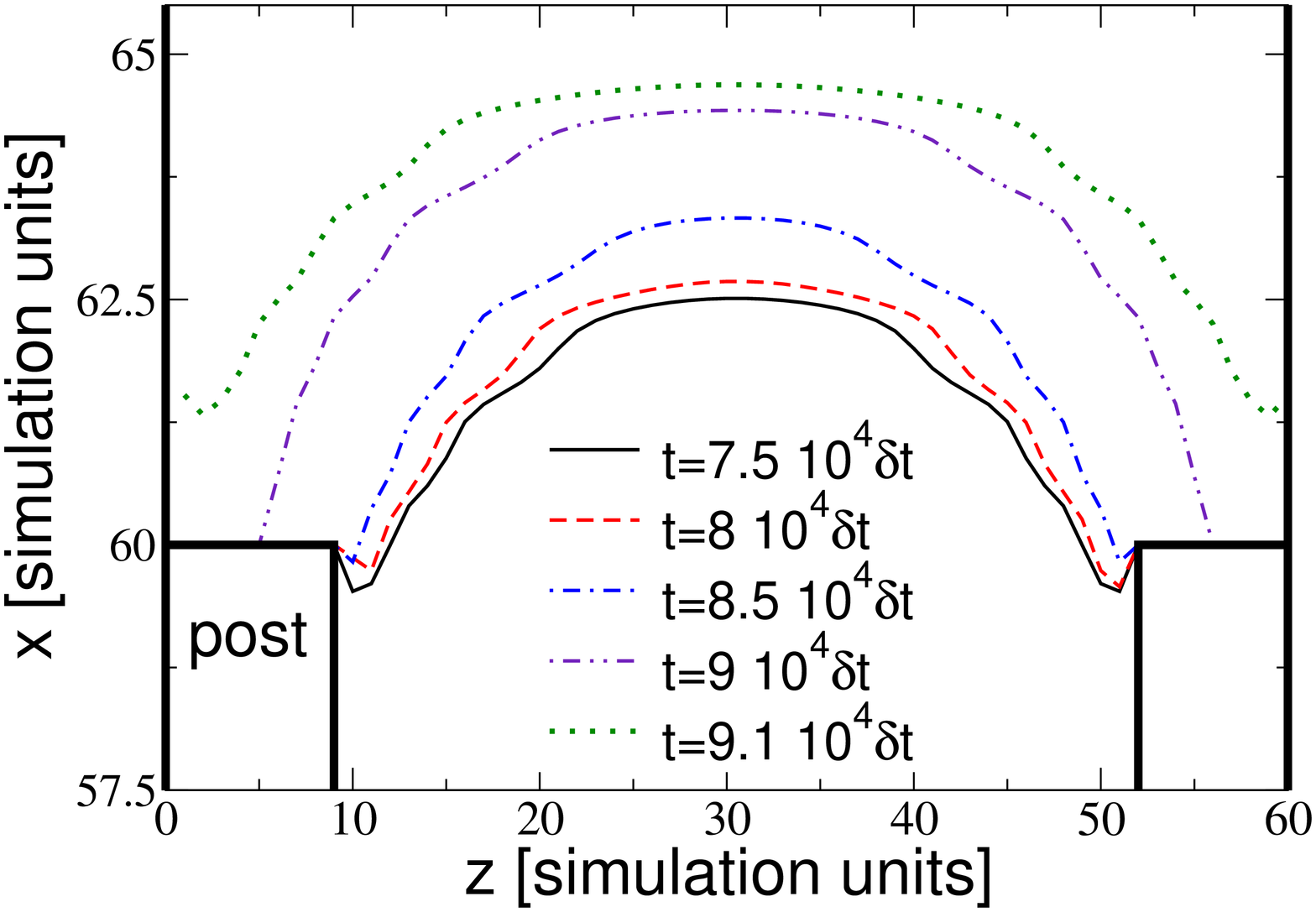}\\
(a) $\qquad\qquad\qquad\qquad\qquad\qquad\qquad\qquad\qquad\quad$ (b)
\caption{(Color online) 
Depinning mechanism for (a) $H=80$, $H_\mathrm{eff}=30$, $L=40$ and 
$\theeq=55^\circ$; (b) $H=60$, $H_\mathrm{eff}=40$, $L=40$
and $\theeq=60^\circ$.
In the first row we report three dimensional interface profiles
during depinning, while in the second row the position
of the advancing front along a  section taken through the centre of a 
post at different times. 
In (b) the face of the post is wet both from its ends and sides.}\label{h4_t55} 
\end{figure}
At small values of $H_\mathrm{eff}$ (for which $\theeq=60^\circ$ does not fill) 
the depinning route is the same as for $H_\mathrm{eff}=0$. This is shown in the 
first column of Fig.\ \ref{h4_t55}. As the meniscus advances near the walls of 
the channel the interface between the post remains almost completely flat. The 
posts are wet, as before, from the channel walls towards the centre. This is not 
the case for larger value of $H_\mathrm{eff}$  (full symbols in Fig.\ 
\ref{HEff}). 
In Fig.\ \ref{h4_t55} column (b) the advancing 
front does not remain pinned in the 
middle  of the channel but the post are wet from their ends and sides
 towards the walls.

\subsection{Inertial effects}\label{secIn}

In performing the simulations we found that inertial effects made it difficult 
to determine the exact contact angle at which depinning occurs. If the front 
reaches the free energy minimum which corresponds to pinning with a residual 
kinetic energy it can overshoot, and hence depin, and it is not possible to 
entirely eliminate this effect without prohibitively long simulations. Indeed 
in a physical system the front will approach the posts with a finite velocity and 
whether it will pin will be a balance between $\theeq$ and the extent to which 
the front has been slowed by the viscous drag in the channel.

\begin{figure}[h]
\includegraphics[angle=-90,scale=0.5]{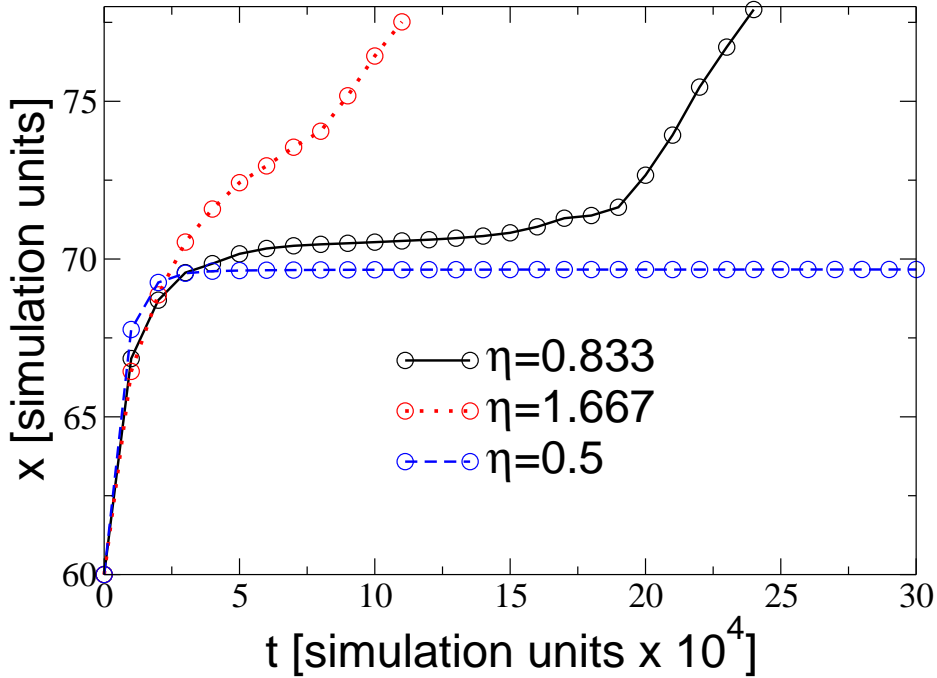}
\caption{Position of the maximum of the advancing front near the wall ($z=2$)
as a function of time
for $D=20$, $L=40$, $\theeq=55^\circ$ and three different values 
of the viscosity $\eta$.
}\label{visc}
\end{figure}

To demonstrate the effects of inertia close to the depinning transition we 
investigated filling a channel with $D=20, L=40, H=60, H_\mathrm{eff}=0 $, 
$\theeq=55^\circ$, from the starting configuration shown in 
Fig.\ \ref{geometry}, for
 three different liquid viscosities. 
Fig.\ \ref{visc} shows the position of the advancing front as a function of
time for each case.
 At early times lower viscosities give slightly higher speeds.
However, the higher viscosity fluids gain more energy from the walls 
and have sufficient inertia to move past the 
pinning position  for this contact angle whereas, at least on the timescale of 
the simulation, the lowest viscosity fluid remains pinned. 
As anticipated in Sec.\ \ref{secHeff0}, inertial effects are also responsible 
for the dependence of $\thetr$ on $L$, which controls the amount of
 water advancing between two posts (or equivalently the scale of the system).

We tried to better determine the position of  $\thetr$ repeating
simulations in which $\theeq$ was gradually decreased (by $0.5^\circ$ each 
$10^5$ time steps) to better 
reproduce a quasi-static relaxation of the interface. For a channel with 
$L=30$ we obtained  $\thetr = 52.5^\circ \pm 0.5^\circ$. 
 However this estimate is very difficult because of 
the flatness of the free energy profile and the very slow interface velocities.

We stress that, for angles near depinning, inertial effects depend 
primarily on the way in which the interface is pulled beyond the end
of the posts by the surface, but not on its initial position within the 
microchannel. Hence we expect that this is not just an artifact of the
simulations, but that similar inertial effects will occur in 
experimental systems.

\section{Discussion}\label{secDis}

As microfabrication techniques become  standard it is becoming 
possible to design microchannels with complicated internal geometries that 
may prove useful in controlling fluid behaviour. 
As a step towards understanding how fluids move in such channels we have 
investigated capillary filling in microchannels patterned by regularly spaced 
square posts. A consequence of the Gibbs' criterion is that ridges that face 
each other across a channel will always pin a slowly moving interface. We 
show that, if the ridges are replaced by posts, the interface is able to depin 
for sufficiently small contact angles. This is because the meniscus can advance
 along the surfaces of the channel between the posts, thus allowing depinning 
to occur.

For posts which span the channel and for a ratio of channel height to distance 
between the posts $H/L\gtrsim 1$ the depinning threshold $\thetr$ is 
independent of $H/L$ because the two surfaces of the channel act 
independently. $\thetr$ lies between $55^\circ$ and $60^\circ$ and the posts 
are wet from the surfaces towards the centre of the channel. For $H/L\lesssim 
1$ the 
threshold contact angle increases with decreasing $H/L$ as the surfaces act 
cooperatively to reduce interface curvature across the channel. Here the 
posts primarily wet from their sides to their centres.

In the general case in which the posts on opposing sides of the channel 
 are separated by a distance $H_\mathrm{eff}$ two regimes are
present for $H/L\gtrsim 1$. 
At low $H_\mathrm{eff}$ the filling/pinned transition is similar to that for
 $H_\mathrm{eff}=0$, with a threshold value of the equilibrium contact
angle around $\thetr$ and posts which  wet from 
the wall to the centre of the channel. For high enough $H_\mathrm{eff}$,
however, the threshold equilibrium contact angle for depinning increases, 
and the posts can also wet from the centre of the channel towards the walls 
during depinning. These results are in agreement with 
 \cite{Kusumaatmaja08}, where channels with fixed $H$
and $H_\mathrm{eff}$ and several $L$ were considered. In particular in 
the range $H/L \gtrsim 1$, a contact angle threshold compatible with
$\thetr$  was observed. On 
increasing the value of $L$ (i.e.\ exploring the $H/L\lesssim 1$ regime),
filling at higher $\theeq$ was found, in agreement with our 
$H_\mathrm{eff}=0$ results.

We have concentrated mainly on the quasi-static situation where inertia is 
neglected and therefore our threshold values are relevant to a very slowly 
moving interface. We have, however, shown that close to the threshold even 
tiny interface velocities can aid depinning. It will now be interesting to 
investigate the more general case of a moving interface and assess the extent 
to which dissipation at the posts can slow and eventually pin the interface 
for a range of contact angles.

The contact angle of a fluid within a microchannel can rather easily be varied 
by applying an electrowetting potential. This opens the possibility of 
controlling the fluid motion by switching $\theeq$ in and out of the pinning 
regime \cite{PrepEW}. This is of particular interest in the large $H/L$ regime 
because the threshold equilibrium contact angle is well approximated by 
$\thetr$, independent of the channel geometry.
\\ \\
{\bf ACKNOWLEDGMENTS}\\
We thank M.\ Blow, H.\ Kusumaatmaja and O.\ Pierre-Louis for useful discussions.
Financial support through the EU project INFLUS is acknowledged.

\end{document}